\documentclass[11pt]{article}

\usepackage{amsmath,mathtools,amssymb,color,epsfig,cite}
\usepackage{amsfonts}
\usepackage{graphicx}
\usepackage{set	space}
\usepackage{mathrsfs}
\usepackage{tikz}
\usepackage{makecell}
\usepackage{multirow}
\usepackage{enumitem}
\usepackage{float}

\makeatletter
\newsavebox{\@brx}
\newcommand{\llangle}[1][]{\savebox{\@brx}{\(\m@th{#1\langle}\)}
  \mathopen{\copy\@brx\kern-0.5\wd\@brx\usebox{\@brx}}}
\newcommand{\rrangle}[1][]{\savebox{\@brx}{\(\m@th{#1\rangle}\)}
  \mathclose{\copy\@brx\kern-0.5\wd\@brx\usebox{\@brx}}}
\makeatother

\usepackage{amsfonts}

\newcommand{\hoch}[1]{$\, ^{#1}$}

\makeatletter
\@addtoreset{equation}{section}
\makeatother

\newcommand{\be}{\begin{equation}}
\newcommand{\ee}{\end{equation}}
\newcommand{\bea}{\setlength\arraycolsep{2pt} \begin{eqnarray}}
\newcommand{\eea}{\end{eqnarray}}

\def\biblio{\bibliographystyle{utphys}\bibliography{LambdaPhi4}}

\def\0{{\sst{(0)}}}
\def\1{{\sst{(1)}}}
\def\2{{\sst{(2)}}}
\def\3{{\sst{(3)}}}
\def\4{{\sst{(4)}}}
\def\5{{\sst{(5)}}}
\def\6{{\sst{(6)}}}
\def\7{{\sst{(7)}}}
\def\8{{\sst{(8)}}}
\def\sst#1{{\scriptscriptstyle #1}}

\def\G{\mathcal{G}}

\def\E{{\mathbb E}}

\def\E{\scriptscriptstyle{E}}
\def\G{\scriptscriptstyle{G}}
\def\0{\scriptscriptstyle{0}}
\def\1{\scriptscriptstyle{1}}
\def\p1{\scriptscriptstyle{(1)}}
\def\2{\scriptscriptstyle{2}}
\def\pp2{\scriptscriptstyle{(2)}}
\def\3{\scriptscriptstyle{3}}
\def\ppp3{\scriptscriptstyle{(3)}}
\def\-{\scriptscriptstyle{-}}
\def\+{\scriptscriptstyle{+}}

\title{$\lambda\,\phi^4$ results - third attempt}

\begin{document}
\def\biblio{}

\vspace*{15pt}
\begin{center}
{\Large {\bf ANEC in $\lambda \, \phi^4$ theory}}

\vspace{35pt}
{\bf Teresa Bautista,\hoch{1,2} Lorenzo Casarin\hoch{2} and Hadi Godazgar\hoch{2}}

\vspace{30pt}

{\hoch{1} \it Department of Mathematics, King's College London\\
The Strand, London WC2R 2LS, UK\\
\hoch{2} \it Max-Planck-Institut f\"ur Gravitationsphysik (Albert-Einstein-Institut) \\
M\"uhlenberg 1, D-14476 Potsdam, Germany
}

 \vspace{35pt}
December 9, 2020

\vspace{45pt}

\underline{ABSTRACT}
\end{center}

\noindent
    Motivated by the goal of applying the average null energy condition (ANEC) to renormalisation group flows, we calculate in \(\lambda \phi^4\) theory the expectation value of the ANEC operator in a particular scalar state perturbatively up to third order in the quartic coupling and verify the expected CFT answer. The work provides the technical tools for studying the expectation value of the ANEC operator in more interesting states, for example tensorial states relevant to the Hofman-Maldacena collider bounds, away from critical points.   
 \noindent

\thispagestyle{empty}

\vfill
{\it Emails:
teresa.bautista@kcl.ac.uk, lorenzo.casarin@aei.mpg.de,  hadi.godazgar@aei.mpg.de}

\pagebreak
\tableofcontents
\vspace{10mm}
\rule{400pt}{0.03pt}
\section{Introduction}

Quantum field theories are our most developed framework for understanding a range of phenomena from particle physics to many-body systems. However despite their success there are many fundamental questions that are still beyond our technical means: the most fundamental question concerning the very existence and definition of quantum field theories non-perturbatively.

However, we are encouraged by recent progress in quantum field theories with conformal symmetry, conformal field theories, where the numerical, and more recently, analytic bootstrap programmes have provided deep insights \cite{Poland:2018epd}. A crucial ingredient in the analytic bootstrap programme~\cite{Komargodski:2012ek, Fitzpatrick:2012yx, Li:2017lmh, Costa:2017twz, Caron-Huot:2017vep, Caron-Huot:2020adz, Carmi:2020ekr} has been the use of causality \cite{Hartman:2015lfa} which requires a Lorentzian perspective.

Meanwhile, the bootstrap approach has also been applied in cosmology \cite{Sleight:2019hfp, Baumann:2019oyu, Baumann:2020dch}, which motivates momentum space considerations \cite{Maldacena:2011nz, Creminelli:2012ed} for CFT correlators generally \cite{Coriano:2013jba, BMS:imp, BMS:renoms, Coriano:2017mux, BMS:renomtj, BMS:renomsjt, Coriano:2018bsy, Bzowski:2019kwd, Coriano:2019nkw, Coriano:2020ees, Jain:2020rmw, Bzowski:2020kfw} which now also have Lorentzian analogues \cite{Bautista:2019qxj, Gillioz:2018mto}, with applications \cite{Gillioz:2020mdd}. 

Indeed the utility of considering theories in Lorentzian signature is being appreciated in a wide range of subjects, for example conformal truncations \cite{Katz:2016hxp, Anand:2019lkt} and most importantly for our purposes, average energy conditions \cite{HofMal, Hartman:2016lgu}---in fact there is also a novel approach to CFTs using more general null-integrated operators \cite{Kravchuk:2018htv, Kologlu:2019mfz}.  
Lorentzian space methods will also be crucial for an understanding of the analyticity property of CFT correlators \cite{Lorce:2020bsg, Maharana:2020fas} more generally.

In this paper we are interested in the average null energy condition (ANEC). The null energy condition states that the components of the energy-momentum tensor along null curves is non-negative,
\begin{equation}
T_{\mu \nu} k^{\mu} k^{\nu} \geq 0
\end{equation}
for some null vector field $k.$ Physically this is the statement that a null observer will observe non-negative energy density. This is violated in the quantum theory by, for example, the Casimir effect. The ANEC is the statement that the integral of the null energy over the complete null worldline is non-negative,
\begin{equation}\label{ANEC-definition}
\mathcal{E}(x^{\+},\hat x):=\int dx^- \,T_{--}(x) \geq 0.
\end{equation}
Intuitively, this means that any violation of the null energy condition is such that it is rendered negligible in an average over the whole worldline.

While it is straightforward to show that the ANEC is satisfied in free theory \cite{Klinkhammer:1991ki}, within the last few years it has been shown to hold for interacting unitary QFTs with a nontrivial UV fixed point using field-theoretic methods \cite{Hartman:2016lgu} and more generally for any unitary QFT using entropy arguments \cite{Faulkner:2016mzt}. This is a rare example of a constraint that is satisfied by a wide class of quantum field theories. Furthermore, already in the case of conformal field theories, the ANEC implies nontrivial bounds, the Hofman-Maldacena bounds, on the $d=4$ conformal anomaly coefficients $a$ and $c$ \cite{HofMal}, which have also been shown to hold by general CFT methods independent of ANEC \cite{Hofman:2016awc}. These bounds apply to any unitary CFT, demonstrating the power of such arguments.

We are, therefore, motivated to understand the consequences of the ANEC in general QFTs. In particular, we wish to understand its implications for renormalisation group flows and discover what the analogues of the Hofman-Maldacena bounds are away from critical points. Such an understanding could lead to insights on the $a$-theorem \cite{Komargodski:2011vj} providing, for example, an interpolating function in terms of the 3-point function of energy-momentum tensors.

In this manuscript, we initiate this study with the more modest goal of understanding the ANEC in the particular example of $\lambda \phi^4$ theory. This theory has the advantage of being simple enough to explore the expectation value of the ANEC operator in explicit detail, while also being an interacting theory with a trivial fixed point in $d=4$ dimensions and a nontrivial, Wilson-Fisher, fixed point in $d = 4 - 2 \epsilon$ dimensions.  Furthermore, given that the field-theoretic arguments of \cite{Hartman:2016lgu} ought not apply to this example, since, rather than a nontrivial UV fixed point, the theory has in fact a Landau pole, this example may also give clues on how to generalise the result of \cite{Hartman:2016lgu} to include the wider class of theories for which the ANEC has been shown to hold \cite{Faulkner:2016mzt}.

More concretely, we evaluate the energy flux on a state corresponding to a single scalar field up to third order in $\lambda$, following the Hofman-Maldacena prescription \cite{HofMal}. According to this precription, the \emph{energy flux} is the nonnormalised expectation value of the ANEC operator inserted at null infinity.\footnote{In the above, and the abstract, we have made no distinction between the ANEC and the energy flux (ANEC at null infinity) operators.} Its positivity then follows from the ANEC. 

To compute it, we first find the Euclidean correlation function of a 
single energy-momentum tensor and two scalar fields, and then use the method given in \cite{Bautista:2019qxj} to find the momentum-space Lorentzian correlator. As advocated in \cite{Bautista:2019qxj}, the Hofman-Maldacena prescription is both conceptually clearer and technically easier to implement in momentum space and this is also confirmed in its application to $\lambda\phi^4$ theory. 

Technically, we find that, despite the complicated form of the 3-point Wightman function, the energy flux relies on only a few contributions. In particular, even though we find a correction at order $\lambda^3$ to the energy flux, this is exactly canceled by the same correction coming from the norm of the state, so that the physical quantity measured by a calorimeter, the \emph{normalised energy flux}, is not corrected and is equal to the free theory result as expected from rotational symmetry.\footnote{We would like to thank the referee for explanations on this point.}

 ANEC expectation values on scalar states, such as the ones considered in this work, are much simpler than those on tensorial states created for instance by the energy-momentum tensor itself. Indeed, as in the case of a CFT, positivity of the energy flux from scalar states does not yield bounds on conformal anomalies, and trivially follows from the positivity of the energy of the state. Nevertheless, the simplicity of these states precisely offers a useful setting to develop the technical tools needed to study the energy flux in more complicated tensorial states in a general QFT. In particular, we will apply these techniques to explore the expectation value on a state created by the energy-momentum tensor in a separate work. Such an investigation will effectively give the Hofman-Maldacena bounds for the conformal anomalies away from the fixed points in the particular example of $\lambda \phi^4$ theory and thereby bring us closer to our goal.

In section \ref{sec:Gaussian}, we set conventions and describe, in Euclidean signature, the correlation functions that we are interested in this work, namely the 3-point function of two scalar field insertions and the energy-momentum tensor, and the 2-point function of the scalar field. In section \ref{sec:Wick-rot}, we describe the Wick rotation needed to find the Wightman functions, which we use, in section \ref{sec:ANEC1} to evaluate the energy flux at tree level in a Gaussian state. The Gaussian state is required in order to regularise distributions that are not well defined. However, beyond tree level the evaluation of the energy flux in the Gaussian state is unwieldy. Hence in section \ref{sec:positive}, we use a different state, which we call the positive-energy state, and calculate the energy flux up to third order in $\lambda$. Finally in section \ref{sec:energy}, we confirm that the normalised energy flux is indeed simply the energy of the state.  

\section{Gaussian-localised state}\label{sec:Gaussian}

In this section we compute the energy flux at tree level. 
We are interested in such an energy flux in one of the simplest states: that created by a single insertion of the operator $\phi$. The energy flux follows from the Wightman 3-point function $\langle\phi \,T_{\mu\nu}\,\phi\rangle$, which we will obtain by first computing the Euclidean 3-point function $\langle\phi \,T_{\mu\nu}\,\phi\rangle_{\scriptscriptstyle{E}}$ and then doing a Wick rotation.\footnote{We use the Euclidean metric \(\delta_{\mu\nu}=(++++)  \), the Lorentzian \(\eta_{\mu\nu}=(-+++)\). Coordinates are \(x^\mu\), light cone coordinates are \(x^{\pm} = x^0 \pm x^1\) and we denote the \(d-2\) transverse components \(x^\mu\), \(\mu\neq 0,1\) with \(\hat x\).} 
The Euclidean $\lambda \phi^4$ action in flat space reads
\begin{equation}
S=\int d^dx \,\left(\frac{1}{2}\,(\partial\phi)^2 +\frac{\lambda}{4!}\,\phi^4\right),
\end{equation}
from which follows the Euclidean energy-momentum tensor 
\begin{equation}\label{EMtensor}
T_{\mu\nu}=\partial_\mu\phi\,\partial_\nu\phi -\frac{1}{2} \,(\partial\phi)^2\,\delta_{\mu\nu}  -\frac{\lambda}{4!}\,\phi^4\,\delta_{\mu\nu} -\xi \left(\partial_\mu\partial_\nu -\delta_{\mu\nu}\,\partial^2\right) \phi^2.
\end{equation}
The last term is the improvement term, which can be seen as following from the addition of $\frac{\xi}{2}\,R_g\,\phi^2$ to the action on a general background spacetime, where $R_g$ is the Ricci scalar of the background metric $g$. At the critical value \(\xi = \frac{d-2}{4(d-1)}\), this additional term makes the energy-momentum tensor traceless on-shell, in agreement with the conformal symmetry of the theory on flat spacetime.
 Moreover, the addition of the \(\xi\)-term makes the energy-momentum tensor renormalisable at the quantum level \cite{Freedman:1974ze}.

We will do the computation in momentum space. We will use the standard double bracket notation for the correlator from which the momentum conserving $\delta$-function has been removed:
\be
\langle \phi(p_1)\,T_{\mu\nu}(p_2)\,\phi(p_3)\rangle=(2\pi)^d\,\delta^{(d)}(p_1+p_2+p_3)\,\llangle \phi(p_1)\,T_{\mu\nu}(p_2)\,\phi(p_3)\rrangle.
\ee
In the double-bracketed correlator $p_3$ is understood to be $p_3=-p_1-p_2$. We use this double bracket notation in both Euclidean and Lorentzian signatures.

The first step is then to compute the  Euclidean 3-point function. We do this with 
the usual perturbative evaluation of correlators with the path integral method. 
A simplification can already be made from the beginning: terms in \eqref{EMtensor} proportional to $\delta_{\mu\nu}$ become proportional to $\eta_{\mu\nu}$ in Lorentzian signature, and since $\eta_{--}=0$, they do not contribute to the ANEC operator. We can similarly drop all terms proportional to $\delta_{\mu\nu}$ that arise from the evaluation of the 3-point function. The resulting 3-point function at tree level reads
\begin{equation}\label{3pnt-E-tree-full}
\llangle \phi(p_1)\,T_{\mu\nu}(p_2)\,\phi(p_3)\rrangle_{\scriptscriptstyle{E},\0}=\frac{-2\, p_{1(\mu} \,p_{3\nu)}+2\,\xi\,p_{2\mu}\,p_{2\nu}}{p_1^2\,p_3^2},
\end{equation}
where the subscripts indicate Euclidean and tree level, and the equal sign has to be understood up to terms proportional to $\delta_{\mu\nu}$.

To compute the normalised energy flux we need to normalise it with the norm of the state, which follows from the Wightman 2-point function of the scalar field. We similarly compute the latter from the Wick rotation of the Euclidean 2-point function. At tree level and in momentum space, this 2-point function can be normalised as
 \be \label{E2pt}
 \llangle \phi(p)\,\phi(-p)\rrangle_{\E,\0}=\frac{1}{p^2},
 \ee
 where again the double bracket notation indicates that the $\delta$-function corresponding to momentum conservation has been removed, i.e.
 \be
 \langle \phi(p)\,\phi(q)\rangle=(2\pi)^d\,\delta^{(d)}(p+q)\,\llangle \phi(p)\,\phi(-p)\rrangle.
 \ee

 \subsection{Wick rotation}\label{sec:Wick-rot}
 
To compute norms and expectation values in Lorentzian signature we need to compute Wightman 2- and 3-point functions. In position space, Wightman functions follow easily from Euclidean correlation functions by means of the usual Wick rotation $t_{\scriptscriptstyle{E}} = i t$ together with the $i\epsilon$ prescription \cite{Haag:1992hx} (see \cite{Hartman:2015lfa} for a review): given any two insertions in the correlator, the time component of the operator to the left acquires a more negative imaginary part than that of the operator to the right.

 In momentum space, there is no such simple prescription to compute Wightman functions from Euclidean correlators. Indeed, it is well known that analyticity in momentum space has to encode the many different causality relations between insertion points, which is achieved, in particular, with Heaviside step functions and Dirac $\delta$-functions, that cannot be easily prescribed in a standard manner. Alternatively, one can Fourier transform the Lorentzian position space expressions, but such transforms become cumbersome in general spacetime dimensions and for correlation functions with more than two insertions. 
 
 In \cite{Bautista:2019qxj}, a method was proposed to compute Wightman functions in momentum space by performing the Wick rotation inside the Fourier transform. 
 The advantage of this method, as opposed to Fourier transforming, is that no complicated integrals need to be performed, and only the analyticity properties of the Euclidean correlation function in momentum space play a role. The method was shown to be practical in computing conformal 3-point correlators. We will hence use this method to compute the Wightman functions that are of interest. For more details see \cite{Bautista:2019qxj}.

With this method, the tree-level Lorentzian 2-point function of the scalar field is
\be\label{2pnt-L-tree}
\llangle \phi(p) \phi(-p)\rrangle_{\0}=	\int d{^d x}\ e^{ipx}
	\langle{\phi(x)\phi(0)}\rangle_{\0}
={} 
2\pi \,\frac{\delta(p^{\0}-|\vec{p}\,|)}{p^{\0}+|\vec{p}\,|}.
\ee

Wick rotation of 3-point functions, whether scalar or tensorial, can be obtained using the same method. The Euclidean 3-point function we need to Wick rotate $\llangle \phi(p_1)\,T_{\mu\nu}(p_2)\,\phi(p_3)\rrangle_{\scriptscriptstyle{E}}$ contains tensorial factors of the type $p_{j\mu}\,p_{k\nu}$.  Given that we only need the null-null component of the 3-point function, $\llangle \phi(p_1)\,T_{--}(p_2)\,\phi(p_3)\rrangle$, it is convenient to directly Wick rotate this component. This can be done by simply writing the null components of the momentum vectors in the Euclidean 3-point function as
\be
p_{\-}=-\frac{p^{\+}}{2}=\frac{i\,p_{\E}-p^{\1}}{2}.
\ee
We will henceforth directly consider this component of the 3-point function, understanding the null direction in the Euclidean signature as given by the complex combination above. 
 
The definition of the ANEC operator entails now another important simplification: the integral over the null coordinate $x^{\-}$ translates, in momentum space, into the vanishing of the $p^{\+}$ component of the momentum:
 \begin{equation}
 \mathcal{E}(x^{\+},\hat x):=\int dx^- \,T_{--}(x) \qquad \rightarrow\qquad
 \mathcal{E}(p^{\-},\hat p) = T_{--}(p^{\+}=0,p^{\-},\hat p),
 \end{equation}
 hence terms proportional to $p_{\-} (=-p^{\+}/2)$ in the energy-momentum tensor do not contribute to the ANEC operator. 
 Similarly, terms in the Euclidean 3-point function $\llangle\phi(p_1)\,T_{\mu\nu}(p_2)\,\phi(p_3)\rrangle_{\scriptscriptstyle{E}}$
proportional to $p_{2\mu}$ can be neglected.\footnote{The Euclidean $p_{-}$ as a tensorial component is simply Wick-rotated to the Lorentzian $p_{-} = - \frac{p^{\0} + p^{\1}}{2}.$ } Effectively, we are interested in evaluating the 3-point function in a momentum configuration with \(p_2{\-} =0\), or equivalently with \(p_1{\-} =-p_3{\-}\), by momentum conservation. We will make use of this property in the following.

The Lorentzian tree-level 3-point function contributing to the energy flux is thus
\begin{equation}\label{3pnt-L-tree}
\llangle \phi(p_1)\,T_{--}(p_2)\,\phi(p_3)\rrangle_{\0} = 4\pi^2\, (p_1^{\+})^2\, \frac{\delta(p_1^{\scriptscriptstyle{0}}-|\vec{p}_1|)}{p_1^{\scriptscriptstyle{0}}+ |\vec{p}_1|}\,\,\frac{\delta(p_3^{\scriptscriptstyle{0}}+|\vec{p}_3|)}{-p_3^{\scriptscriptstyle{0}}+|\vec{p}_3|}.
\end{equation}
 
 Notice that as opposed to the energy-momentum tensor \eqref{EMtensor}, the ANEC operator does not depend on the improvement parameter $\xi$. Since $\xi$ does not show up in the action either, the energy flux does not depend on this parameter at all.

 \subsection{Energy flux}
 \label{sec:ANEC1}
 
 We are interested in computing the expectation value of the ANEC operator $\langle\mathcal{E}\rangle$ on momentum eigenstates 
 \begin{equation}\label{wave-packet}
|\phi(q)\rangle\equiv\int d^d x\, e^{-i q\, x^{\0}} \,\phi(x)|0\rangle
 \end{equation}
 at null infinity. 
 The reason is that when the ANEC operator is inserted at null infinity, it commutes with the momentum operator. In this case, these states become eigenstates of the ANEC operator, and the ANEC induces optimal constraints from the resulting energy flux.

In \cite{HofMal}, Hofman and Maldacena use such states to put optimal constraints on conformal anomalies, the conformal collider bounds. They compute the energy flux in conformal field theories, using states not only created by scalar operators but also by currents and the energy-momentum tensor. The energy flux follows from the corresponding conformal 3-point functions, which in the latter case is the correlator of three energy-momentum tensor insertions, which depends the on conformal anomaly coefficients $a$ and $c$. Thus, non-negativity of the energy flux implies inequalities between conformal anomaly coefficients. 

Being a bit more concrete, the conformal collider bounds actually follow from imposing the non-negativity of the time-integrated normalised energy flux $E$ as measured by some calorimeter placed at null infinity. This quantity must be positive since it is an energy measurement, and turns out to be determined by the expectation value of the ANEC operator placed on the sphere at null infinity,
\begin{equation}\label{ANEC-position-noGaus}
\langle\mathcal{E}(q)\rangle=\lim_{r\rightarrow\infty}\, r^{d-2}\, \langle \phi(q)^\dagger\,\int\limits_{-\infty}^\infty dx^-\,T_{--} (x^+,x^-)\,\phi(q)\rangle ,
\end{equation}
where $r=x^+/2$ in this limit and the expectation value is taken on the states $|\phi(q)\rangle$. The energy-momentum tensor, which represents the calorimeter, is inserted at $(x^+\rightarrow\infty,x^-,\hat x=0)$. The $r^{d-2}$ prefactor is required in order for this quantity to be an energy flux and is, moreover, crucial for obtaining a finite result when the limit is taken. 
Finally, the actual normalised energy flux measurement follows from normalising this expectation value with the norm of the state, which is given by the Wightman 2-point function of the operator creating the state. The calorimeter hence measures
\be\label{aab}
E=\frac{\langle\mathcal{E}(q)\rangle}{
 \langle 
	\phi(q)| \phi(q) 
 \rangle }
 \geq 0\,.
\ee

 In a conformal field theory, 3-point functions are fully fixed by conformal symmetry up to constants. In particular, the 3-point function of two scalar operators and the energy-momentum tensor has only one free constant, which is fixed with the normalisation of the scalar operator 2-point function through Ward identities. 
Indeed the ANEC on scalar states simply measures the energy of the state,
 \be\label{Energy-measurement-conf}
 E=\frac{q}{S_{d-2}},
 \ee
 where $S_{d-2}$ is the surface area of the $(d-2)$-dimensional sphere and it is assumed that $q\geq0$.  More generally, rotational symmetry guarantees the same answer, namely equation \eqref{Energy-measurement-conf} holds in any QFT, where the state is a momentum eigenstate.  Its positivity therefore is equivalent to the positivity of the energy of the state.
 
 We are interested in the implications of the ANEC away from conformality.  Our goal is to use the scalar state to compute the energy flux in $\lambda \phi^4$ theory, verifying \eqref{Energy-measurement-conf}. We use this simpler
calculation to develop our methods and gain technical intuition. These will be required in the more complicated but interesting cases of expectation values on tensorial states, which we leave for future work.
 
Rather than following the position space approach of \cite{HofMal}, we evaluate the energy flux in momentum space. This is both because the diagrammatic computation of the 3-point functions is simpler in momentum space, and because the state in which we evaluate the expectation value is a momentum eigenstate. Indeed, as explained in \cite{Bautista:2019qxj}, momentum space simplifies the computation and highlights the role of the limit to null infinity, $r\rightarrow\infty$, in the interpretation of this quantity as an expectation value.
 To reproduce the expected conformal results, we also chose a state with purely-timelike momentum $(q,\vec{0})$.
 
In \cite{HofMal}, given a state defined in equation \eqref{wave-packet}, Hofman and Maldacena identify \eqref{ANEC-position-noGaus} with
\begin{equation}\label{aaa}
\langle\mathcal{E}(q)\rangle=
\lim_{r\rightarrow\infty}\, r^{d-2}\,\int\limits_{-\infty}^\infty dx^- \int  d^d y \ e^{iq y^0}\langle \phi(y) \, T_{--} (x^+,x^-,\hat x =0)\,\phi(0)\rangle
\end{equation}
where a (formally divergent) factor  of the spacetime volume $\rm V$ is dropped, since it is canceled by the same factor in the denominator. Indeed, the norm of the state is related to the 2-point function \eqref{2pnt-L-tree} via 
\be \label{norm}
\langle 
\phi(q)| \phi(q) 
\rangle = {\rm V}   \llangle 
\phi(q)\phi(-q) 
\rrangle.
\ee

Equation \eqref{aaa} can also be derived using states defined as insertions in imaginary time \cite{Hartman:2016lgu}, which is indeed how we also define states in this paper. 

 In momentum space equation \eqref{aaa} becomes
 \begin{align}\label{ANEC-momentum-noGaus}
\langle \mathcal{E}(q)\rangle=
2\,\lim_{r\rightarrow\infty}\, r^{d-2}\,
\int\frac{d^{\scriptscriptstyle{d-1}}\vec{p}}{(2\pi)^{\scriptscriptstyle{d-1}}}\,\,\,e^{ 2i  p^{\1} r}\,
\llangle \phi(q,\vec{0}\,)\,T_{--}(-p^{\1},\vec{p}\,)\,\phi(p^{\1}-q,-\vec{p}\,)\rrangle.
\end{align}
In the ANEC operator, the integral over \(x^-\)  gives  $\delta(p^{\+})$, which we have used to integrate over \(p^{\0}\).

The  energy flux at tree level follows from plugging in the tree-level expression for the 3-point function \eqref{3pnt-L-tree} into the rhs of \eqref{ANEC-momentum-noGaus}:
 \begin{align}\label{aac}
\langle \mathcal{E}(q)\rangle
\propto \mathrm{Vol}(\mathbb{R}^{d-2})\,q^{d-2}\,\delta(q),
\end{align}
which is ill defined. The norm of the state \eqref{wave-packet}, which follows from the 2-point function \eqref{2pnt-L-tree}, is
\be\label{acc2}
\llangle\phi(q)\,\phi(-q)\rrangle_{\0}\propto \frac{\delta(q)}{q}.
\ee
This vanishes for $q\neq0$ and is ill-defined at $q=0$. Therefore, regularisation is required  in order to compute the normalised energy flux $ E$.

As suggested in \cite{HofMal}, one can consider a Gaussian wave packet which localises the state.\footnote{The uncertainties arising in the expectation value on a plane wave state can be understood as coming from the fact that a plane wave spreads over all of space, hence overlapping with the calorimeter or energy-momentum tensor insertion.  For the calorimeter's measurement to be well defined, the state needs to create a localised energy perturbation at some point away from the calorimeter's position, which can be achieved by adding some localising factor to the state such as the Gaussian one. } In particular, we chose a state of the form:
\begin{equation}\label{Gaussian-state}
|\phi(q)\rangle_{\G}\equiv \int d^dx \, e^{-i q x^0}\,e^{-\frac{|\vec{x}|^2}{\sigma^2}}\, \phi(x)\,|0\rangle 
\end{equation}
namely a Gaussian-normalised  state which is only localised in space with width $\sigma$, but not in time.\footnote{Notice that this is slightly different from the Gaussian factors used in \cite{HofMal}, where space and time are treated in the same way. However, in \cite{HofMal} the Gaussian factors do not actually play a role and are dropped in the evaluation of the ANEC.} 
With such a state, in the spirit of  \cite{HofMal}, we generalise \eqref{aaa} to 
\begin{align}
&\langle \mathcal{E}(q)\rangle_{\G}  
{}= \lim_{r\rightarrow\infty}\, r^{d-2}\,\int\limits_{-\infty}^\infty dx^- \int  d^d y \ e^{iq y^0} e^{- \frac{|\vec y|^2}{\sigma^2} }\langle \phi(y) \, T_{--} (x^+,x^-,\hat x =0)\,\phi(0)\rangle
\\ & {}=
\frac{2\sigma^{d-1}}{\pi^{\frac{d-1}{2}}}\int \frac{d^{\scriptscriptstyle{d-1}} \vec{k}}{(2\pi)^{\scriptscriptstyle{d-1}}}\, e^{-\frac{\sigma^2\,\vec{k}^2}{4}}\lim_{\scriptscriptstyle{r\rightarrow\infty}} r^{d-2}
\int\frac{d^{\scriptscriptstyle{d-1}}\vec{p}}{(2\pi)^{\scriptscriptstyle{d-1}}}\,e^{ 2i  p^{\1} r}
\llangle \phi(q,\vec{k}\,)\,T_{--}(-p^{\1},\vec{p}\,)\phi(p^{\1}-q,-\vec{p}-\vec{k}\,)\rrangle
\end{align}
and we normalise  it with 
 \be\label{2pt-L-Gaussian}
 \llangle\phi(q)\,\phi(-q)\rrangle_{\G} \equiv
 	\int d{^d x}\ e^{iq x^{\0}}e^{-\frac{|\vec x|^2 }{\sigma^2}}
 	\langle{\phi(x)\phi(0)}\rangle.
 \ee 
Using the tree-level 3-point function \eqref{3pnt-L-tree}, we now obtain
\be\label{ANEC-tree-Gaussian}
\langle \mathcal{E}(q)\rangle_{\G,\0}= \frac{1}{2^{d-1}\, \sqrt{\pi}^{d-3}}\,\,\sigma^{d-1}\,e^{-\frac{\sigma^2\,q^2}{4}}\,\,q^{d-2}, \qquad \llangle\phi(q)\,\phi(-q)\rrangle_{\G,\0} 
 =\frac{2^{2-d}\,\pi}{ \Gamma(\frac{d-1}{2})}\,\sigma^{d-1}\,e^{-\frac{\sigma^2\,q^2}{4} } \,\,q^{d-3},
 \ee 
where again it is assumed that $q\geq0$ (if $q<0$, $\langle \mathcal{E}\rangle$ vanishes due to a Heaviside step function that has been dropped).
 Taking the limit $ \sigma \, q \rightarrow\infty$ reproduces the ill-defined expressions \eqref{aac} and \eqref{acc2}. 
 The limit is ill-defined because it is only non-trivial for a state with vanishing $q$. This is a particularity of the one-field-insertion state, and would not appear for states created by multiple field insertions or generic CFT operators, which generically have finite norm for finite energy $q$. See below \eqref{Eflux-tree} and appendix \ref{appendix} for further explanations.

The tree-level normalised energy flux now becomes
 \be\label{Eflux-tree}
 E_{\0}=\frac{\langle \mathcal{E}(q)\rangle_{\G,\0}}{\llangle\phi(q)\,\phi(-q)\rrangle_{\G,\0}}=\frac{q}{S_{d-2}},
 \ee
 where we use $S_{d-2}=(4\pi)^{d/2-1}\Gamma(d/2-1)/\Gamma(d-2)$. This flux is non-negative because $q\geq0$, and correctly reproduces \eqref{Energy-measurement-conf}. Hence, the localisation of the Gaussian state effectively `allows' the momentum $q$ to be purely timelike and regularises the energy flux.
 
 The need for the Gaussian regularisation is particular to the one-field-insertion state we have considered. For general scalar states, the $i\epsilon$ prescription employed to compute the Wightman function effectively acts as a regulator for the norm, by giving the Lorentzian time a small Euclidean off-set, yielding a finite norm everywhere. In the case of the one-field insertion state, this regularisation precisely gives a Dirac delta function for the norm, which for our purposes then requires further regularisation. Had we instead considered a state created by multiple $\phi$ insertions, hence by a generic operator with conformal dimension $\Delta$, its norm would be given by $\llangle\mathcal{O}(q)\,\mathcal{O}(-q)\rrangle_{\0}\propto q^{2\Delta-d}$. Analogously, the energy flux would be given by $\langle \mathcal{E}(q)\rangle\propto q^{2\Delta-d+1}$, and the normalised energy flux would not require (Gaussian) regularisation. We do not consider such states in this work though because they would make the diagrammatic computation of the perturbative corrections much harder. See also appendix \ref{appendix} for a further discussion of these issues.
 
We next want to compute QFT corrections to the (non)normalised energy flux. While the Gaussian state is necessary in order to regularise the tree-level contributions, it becomes very cumbersome to work with at higher loops. Indeed, the Gaussian state effectively adds a momentum integral to the ones already coming from the higher-loop contributions, making the computation unwieldy. 
We will therefore consider a plane wave state in the next section to calculate higher loop corrections.

\section{Positive-energy state}
\label{sec:positive}

In this section, we consider a momentum eigenstate (i.e. with no Gaussian localisation)
\begin{equation}\label{wave-packet2}
|\phi(q)\rangle\equiv\int d^d x\, e^{ - i q x^{\0}} \,\phi(x)|0\rangle,
 \end{equation}
with strictly positive energy $q>0$. With such a choice, the relevant correlators at tree level vanish:
\be
\llangle \phi(q,\vec{0}\,)\,T_{--}(-p^{\1},\vec{p}\,)\,\phi(p^{\1}-q,-\vec{p}\,)\rrangle_{\0}\propto q\,\delta(q)\frac{\delta(p^{\1}+|\vec{p}\,|)}{|\vec{p}\,|},\qquad 
\llangle \phi(-q)\,\phi(q)\rrangle_{\0}\propto\frac{\delta(q)}{q}.
\ee
Therefore, the first contribution to the energy flux starts at loop level, and the ratio of the leading loop corrections must reproduce  the conformal or tree-level result \eqref{Energy-measurement-conf}.

We start with the 1-loop corrections. The Euclidean 2-point function of the massless scalar field in $\lambda \phi^4$ theory receives no correction at 1-loop in dimensional regularisation because this would come from a tadpole integral, which can be consistently put to zero. The Euclidean 1-loop contribution to the 3-point function turns out to be proportional to
\begin{equation}\label{3pnt-E-1loop}
\llangle \phi(p_1)\,T_{\mu\nu}(p_2)\,\phi(p_3)\rrangle^{\p1}_{\scriptscriptstyle{E}}\sim p_{2\mu}p_{2\nu}-p_2^2\,\delta_{\mu\nu}.
\end{equation} 
As argued in the previous section, terms proportional to the metric or to the momentum $p_{2\mu}$ of $T_{\mu\nu}$ do not contribute. As a consequence, there are no 1-loop corrections to the nonnormalised flux for a state created by a single field insertion (regardless of its momentum configuration). This is to be expected since at order \(\lambda\) there is no wavefunction renormalisation,  and in the stress tensor only \(\xi\) is renormalised \cite{Freedman:1974ze}, but as we mentioned before \(\xi\)-dependent terms do not contribute to the ANEC.
Therefore, the first contribution to the nonnormalised flux is at  2-loops. 

In the reminder of this section we will compute the energy flux up to  order \(\lambda^3\), and in the next section we will compute the 2-point function and the resulting normalised energy flux to the same order.

\subsection{Corrections to the energy flux}

We compute the 2- and 3-loop corrections to the Euclidean 3-point function with usual functional methods. As argued above, we drop terms proportional to the metric or to $p_{2\mu}$, which greatly simplifies the expressions. 

We start by computing the 2-loop correction and then evaluate the energy flux. 
The 2-loop correction to the 3-point function is given by the contributions of two diagrams with a sunset in each leg, plus that of a third diagram corresponding to a sunset with the insertion of the energy-momentum tensor in one of its lines. 
This last contribution involves an integral of the form
\be\label{3pt-E-2loop-aux-integral}
\llangle\phi(p_1)\,T_{--}(p_2)\,\phi(p_3)\rrangle_{\scriptscriptstyle{E}}^{\scriptscriptstyle{(2)}}\sim
\frac{1}{p_1^2 \,p_3^2}
\int \frac{d^{\scriptscriptstyle{d}} k}{(2\pi)^{\scriptscriptstyle{d}}} \frac{(k^{\+}+p_1^{\+})^2}{(k+p_1)^2\,(k-p_3)^2\,k^{4-d}}.
\ee 
Wick rotating the Euclidean expression, the resulting 2-loop contribution of the Lorentzian 3-point function is
\begin{align}\label{3pnt-L-2loop}
&\llangle \phi(p_1)\,T_{--}(p_2)\,\phi(p_3)\rrangle^{\scriptscriptstyle{(2)}}=\notag\\[7pt]
&C_{2s}(\lambda,d)(p_{1}^{\+})^2
\left[\frac{\dot\delta(p_3)}{(p_1^{\0}+|\vec{p}_1|)^{5-d}}\,\frac{d}{dp_1^{\0}}\left(\frac{\theta(p^{\scriptscriptstyle{0}}_1 - |\vec{p}_1|)}{(p_1^{\0}-|\vec{p}_1|)^{4-d}}\right)+
\frac{\bar\delta(p_1)}{(-p^{\scriptscriptstyle{0}}_3 + |\vec{p}_3|)^{5-d}}\,\,
\frac{d}{dp_3^{\0}}
\left(\frac{\theta(-p^{\scriptscriptstyle{0}}_3 -|\vec{p}_3|) }{(-p^{\scriptscriptstyle{0}}_3 - |\vec{p}_3|)^{4-d}}\right)\right]\notag\\[9pt]
&+C_{2b}(\lambda,d)\,\int \frac{d^{\scriptscriptstyle{d}} k}{(2\pi)^{\scriptscriptstyle{d}}}\,\,\Theta(-k^{\0}-|\vec{k}|)\,\,\frac{(k^{\+}+p_1^{\+})^2}{|k^2|^{2-d/2}}\,\,\left[
 \frac{\bar\delta(k+p_1)\,\bar\delta(k-p_3)}{p_1^2\,p_3^2}\,\,+\,\,\frac{\bar\delta(p_1)\,\dot\delta(p_3)}{(k+p_1)^2\,(k-p_3)^2}\right.\notag\\[7pt]
&\hspace{95mm}
\left.+\,\,\frac{\bar\delta(p_1)\,\bar\delta(k-p_3)}{p_3^2\,(k+p_1)^2}\,\,+\,\,\frac{\dot\delta(p_3)\,\bar\delta(k+p_1)}{p_1^2\,(k-p_3)^2}\right]\notag\\[7pt]
&+C_{2p}(\lambda,d)\,\int  \frac{d^{\scriptscriptstyle{d}} k}{(2\pi)^{\scriptscriptstyle{d}}}\,\frac{(k^{\+}+p_1^{\+})^2}{|k^2|^{2-d/2}}\,\, \left[\bar\delta(p_1)\,\dot\delta(k+p_1)\left(\frac{\bar\delta(k-p_3)}{p_3^2}+\frac{\dot\delta(p_3)}{(k-p_3)^2}\right)\right.\notag\\[7pt]
&\hspace{66mm}\left.+\,\,\dot\delta(p_3)\,\dot\delta(k-p_3)\left(\frac{\bar\delta(p_1)}{(k+p_1)^2}+\frac{\bar\delta(k+p_1)}{p_1^2}\right)\right],
\end{align}
where we use the short-hand notation
\begin{equation}\label{notation-deltas}
\bar\delta(p)=\frac{\delta(p^{\0}-|\vec{p}\,|)}{p^{\0}+|\vec{p}\,|}\, ,\qquad \dot\delta(p)=\frac{\delta(p^{\0}+|\vec{p}\,|)}{-p^{\0}+|\vec{p}\,|}\, .
\end{equation}
The \(C_{2s}\) term corresponds to the contributions of the two diagrams with a sunset on each leg. These are given in terms of explicit derivatives because as distributions they are only well defined if understood as acting by integration by parts  \cite{Freedman:1991tk}.\footnote{Indeed, if we act with the derivative on the factor inside the parenthesis we obtain  $(p_1^{\0}-|\vec{p}_1|)^{5-d}$ in the denominator, which for $3<d\leq 4$ is divergent. 
 If instead the derivatives act by integration by parts the distribution is well defined.}

The other terms arise from the Wick rotation of the auxiliary integral \eqref{3pt-E-2loop-aux-integral}, which has a branchcut due to the $k^{4-d}$  factor in the denominator. In the integral with coefficient $C_{2b}(\lambda,d)$, the $k^{\0}$ integral runs along the branchcut as enforced by the Heaviside step function. The \(C_{2p}\) term  comes from $k^{\0}$ poles lying on the branchcut as indicated by the Dirac $\delta$-functions; these poles originally come from the denominator in the Euclidean expression \eqref{3pt-E-2loop-aux-integral}. These poles are actually excluded from the integration range in the branchcut integrals in the \(C_{2b}\) term, where the Principal Value prescription is implicitly used.

The constants $C_{2s}$, $C_{2b}$ and $C_{2p}$ are $d$-dependent numerical constants of order $\lambda^2$. The only relevant one for us will be
\be
C_{2b}(\lambda,d)=\frac{\lambda^2}{32\,(4\pi)^{\frac{d}{2}-3}}\,\frac{\Gamma(d/2-1)}{\Gamma(d-2)}.
\ee

With the expression for the 3-point function, we next compute the 2-loop correction to the energy flux, 
which follows from inserting  \eqref{3pnt-L-2loop} in the expression for the energy flux of a momentum eigenstate with $(q,\vec{0})$, $q>0$, which we repeat here for convenience: 
\begin{align}\label{ANEC-momentum-noGaus-bis}
\langle \mathcal{E}(q)\rangle=
2\,\lim_{r\rightarrow\infty}\, r^{d-2}\,
\int
\frac{d^{{\scriptscriptstyle{d-1}}}\vec{p}}{(2\pi)^{\scriptscriptstyle{d-1}}}\,\,\,e^{ 2i  p^{\1} r}\,
\llangle \phi(q,\vec{0}\,)\,T_{--}(-p^{\1},\vec{p}\,)\,\phi(p^{\1}-q,-\vec{p}\,)\rrangle.
\end{align}
A huge simplification readily follows since all the terms in \eqref{3pnt-L-2loop} proportional to $\bar\delta(p_1)$ give $\delta(q)$ in the above and hence do not contribute. From the remaining terms in \eqref{3pnt-L-2loop}, the three terms proportional to $\dot\delta(p_3)$ do not contribute either. Indeed, this Dirac $\delta$-function is $\dot\delta(p_3)\sim\delta(p^{\1}-q+|\vec{p}\,|)$ and can be used to integrate the $p^{\1}$ integral of \eqref{ANEC-momentum-noGaus-bis}, 
\be
r^{d-2}\,
\int\frac{d^{{\scriptscriptstyle{d-2}}}\hat{p}}{(2\pi)^{\scriptscriptstyle{d-2}}}\,\,\frac{d p^{\1}}{2\pi}\,\,e^{ 2i  p^{\1} r}\,
\frac{\delta\left(p^{\1}-q+|\vec{p}\,|\right)}{|\vec{p}\,|}\,\, f(p^{\1},\hat{p};q;k),
\ee
where $f$ is a function of the momenta, different for each term, and $k$ is the integrated momentum in the last branchcut and last pole terms in \eqref{3pnt-L-2loop}. We can next  do the change of variables $\hat p:=\hat t/r -q \,\hat {1}$, where $\hat {1}$ is an arbitrary constant unit vector, to compute the rest of the $(d-2)$-dimensional $\hat p$ integrals. After taking the $r\rightarrow\infty$ limit, the result is 
\be
\int \frac{d^{{\scriptscriptstyle{d-2}}}\hat{t}}{(2\pi)^{\scriptscriptstyle{d-2}}}\,\, e^{i \,q \,\hat t\cdot\hat 1}\,\, f(0,  - q \hat 1 ;q;k)\,\propto \,\delta^{(d-2)}(q\,\hat 1)\,\, f(0,-q\,\hat 1;q;k),
\ee
 which vanishes since \(q > 0\). In the case of the last branchcut and last pole contributions, there is a remaining integral over the auxiliary momentum $k$ that can be shown to be finite, so that the result vanishes unambiguously.

Hence, the only contribution to the energy flux comes from the first branchcut (first term of the \(C_{2b}\) term), and gives
\be\label{ANEC-2loop}
\langle\mathcal{E}(q)\rangle_{\2}=\frac{\lambda^2}{12\,(4\pi)^{\frac{3d}{2}-2}}\frac{\Gamma(\frac{d}{2}-1)^2}{\Gamma(\frac{3d}{2}-3)}\frac{1}{q^{9-2d}}.
\ee

It is remarkable how the many simplifications brought about by the nature of the ANEC operator, together with our convenient choice of state, has minimised the computation: from all the very many terms present in the 2-loop Lorentzian 3-point function, only one contributes to the energy flux.

We proceed now to compute the 3-loop correction. 
At this order, the Euclidean 3-point function consists of a contribution equal to the 2-loop one but with different coefficients and powers, and an additional contribution  involving two auxiliary integrals of the type \eqref{3pt-E-2loop-aux-integral} (up to terms proportional to either $\delta_{\mu\nu}$ or $p_{2\mu}$). Dropping all terms containing $\bar\delta(p_1)$ or $\dot\delta(p_3)$, which give vanishing contributions by the arguments above, the resulting 3-loop contribution is
\begin{align}\label{3pnt-L-3loop}
&\llangle \phi(p_1)\,T_{--}(p_2)\,\phi(p_3)\rrangle^{\scriptscriptstyle{(3)}}=
C_{3b}\,\int \frac{d^{\scriptscriptstyle{d}} k}{(2\pi)^{\scriptscriptstyle{d}}} \,\,\Theta(-k^{\0}-|\vec{k}|)\,\,\frac{(k^{\+}+p_1^{\+})^2}{|k^2|^{4-d}}\,\,
\frac{\bar\delta(k+p_1)\,\bar\delta(k-p_3)}{p_1^2\,p_3^2}\notag\\[7pt]
&+C_{3bb} \int \frac{d^{\scriptscriptstyle{d}} k \, d^{\scriptscriptstyle{d}} l}{(2\pi)^{\scriptscriptstyle{2d}}} \,\,\frac{\bar\delta(k+p_1)\,\bar\delta(l-p_3)}{p_1^2\,p_3^2} \,\Bigg[\Theta(-l^{\0}-|\vec{l}\,|)\, \frac{(k^{\+} +p_1^{\+})^2}{|l^2|^{2-d/2}}\,\left(\frac{\dot\delta(l+k-p_3)}{(k-p_3)^2}
+\frac{\dot\delta(k-p_3)}{(l+k-p_3)^2}\right)
\notag\\[7pt]
 &\hspace{55mm}+\Theta(-k^{\0}-|\vec{k}\,|)\frac{(l^{\+} +p_1^{\+})^2}{|k^2|^{2-d/2}}\,\left(\frac{\dot\delta(l+k+p_1)}{(l+p_1)^2}+\frac{\dot\delta(l+p_1)}{(l+k+p_1)^2}\right)\Bigg]
\notag\\
&+C_{3pp}\,\int \frac{d^{\scriptscriptstyle{d}} k \, d^{\scriptscriptstyle{d}} l}{(2\pi)^{\scriptscriptstyle{2d}}}  \,\frac{\bar\delta(k+p_1)\,\bar\delta(k-p_3)}{\,p_1^2\,p_3^2}\,\frac{(k^{\+} +p_1^{\+})^2}{|l^2|^{2-d/2}}\,\left[\bar\delta(l-p_3)\,\dot\delta(l+k-p_3)\right.\notag\\
&\hspace{100mm}\left.+\,\,\bar\delta(l+p_1)\,\dot\delta(l+k+p_1)\right].
\end{align}
The first term is of the same form as the term contributing at 2-loops. The remaining terms above  involve two auxiliary momentum integrals. The terms proportional to $C_{3bb}$ correspond to integrals  along branchcuts, whereas the  \(C_{3pp}\) terms  correspond to the contributions of poles lying on these branchcuts. The constants $C_{3i}$ depend on $d$ and $\lambda$, and are of order $\lambda^3$. The only relevant ones are
\begin{align}
C_{3b}(\lambda,d)&=\frac{\lambda^{3}}{16\,(4 \pi)^{d-3}\,(d-4)} \frac{\Gamma(3-d / 2)^{2}\, \Gamma(d / 2-1)^{4}}{\Gamma(d-2)^{2}\,\Gamma(5-d)\,\Gamma(d-3)},\\
C_{3pp}(\lambda,d)&= \frac{\lambda^{3}}{2^7\,(4 \pi)^{\frac{d}{2}-5}\,(d-4)} \frac{\Gamma(3-d / 2)\, \Gamma(d / 2-1)^{2}}{\Gamma(d-2)\,\Gamma(d/2+1/2)\,\Gamma(1/2-d/2)}.
\end{align}

We next proceed with the computation of the 3-loop correction to the energy flux. It can be seen quite generally that the terms given by coefficient \(C_{3bb}\) 
do not contribute to the energy flux. Specialising to the momentum configuration relevant to \eqref{ANEC-momentum-noGaus-bis}, i.e.\ \(p_1 = (q,\vec 0)\) and \(p_3 = (p^{\1} -q , -\vec p)\) (so that \(p_3^{\1} = -p^{\1}\) ) and shifting $l^{\0} \rightarrow l^{\0} - q,$  $\bar\delta(l-p_3)$ becomes
\begin{align}
 \frac{\Theta(l^{\0} - p^{\1})}{2 \, |l^{\+}|} \, \delta \Big(p^{\1} - \frac{l^{\+} l^{\-} - |\hat{p}|^2 - 2 \,\hat{l} \cdot \hat{p} - |\hat{l} |^2}{2 \, l^{\+}} \Big). 
\end{align}
Therefore, letting $\hat{p} \rightarrow \frac{\hat{t}}{r}$ and integrating over $p^{\1}$ and $\hat{t}$, we get
\begin{equation}
 \lim_{r\rightarrow\infty}\, \int d l^{\-}  \,e^{ i \, l^{\-} r} f(l^{\-}),
 \label{bcint}
\end{equation}
where all Dirac $\delta$-functions have been integrated,\footnote{Note that there are remaining integrals, which are not relevant for the following argument.} which also means that the integral over $l^{\-}$ isn't over the full real line. The properties of the function $f(l^{\-})$ for all four terms is that it decays faster than $1/l^{\-}$ as $l^{\-} \rightarrow\pm \infty$ and it has poles and branch points on the real line. The latter property of the function makes the integral \eqref{bcint} formally divergent (unless they are in the region of integration), which means that the integral must be regularised by moving the poles and branch points above or below the real line. Whichever way the integral is regularised, or if it is finite to begin with, we can use integration by parts to show that the integral \eqref{bcint} is at most $O(1/r),$ hence vanishing in the $r \rightarrow \infty$ limit.

Therefore, the 3-loop correction receives contributions from the first and the last two terms in \eqref{3pnt-L-3loop}, and the energy flux up to this order becomes
\be\label{ANEC-3loop}
\langle\mathcal{E}(q)\rangle_{\3}=\frac{\lambda^2}{12\,(4\pi)^{\frac{3d}{2}-2}}\frac{\Gamma(\frac{d}{2}-1)^2}{\Gamma(\frac{3d}{2}-3)}\frac{1}{q^{9-2d}}
\left(1-\frac{3\,\lambda}{(4\pi)^{\frac{d}{2}}}\frac{4d-10}{4-d}\frac{\Gamma(3-\frac{d}{2})^2\,\Gamma(\frac{d}{2}-1)^2\,\Gamma(\frac{3d}{2}-3)}{\Gamma(d-2)\,\Gamma(5-d)\,\Gamma(2d-4)}\frac{1}{q^{4-d}}\right).
\ee
Letting \(d\to 4\), the energy flux is, after renormalization,\footnote{
		We used minimal subtraction and redefined \(M\) to absorb (\(q\)-independent) numerical constants.
	}
\be
\langle\mathcal{E}(q)\rangle_{\3}=\frac{\lambda^2}{24\,(4\pi)^4}\,\frac{1}{q}\,\left(1+\frac{3\,\lambda}{16\,\pi^2}\,\log\frac{q^2}{M^2}\right),
\ee
where $\lambda=\lambda(M)$ is the renormalised coupling at scale $M$. 
This $\lambda$ correction is compatible with the running of the coupling as expected, in the sense that the energy flux only depends on the coupling renormalised at scale $q$.

The corrected energy flux is positive, consistent with the ANEC. 
Indeed, the $\lambda$ correction would only render it negative at energy scales $q$ such that $\log q/M\lesssim -8\pi^2/3\lambda$, where perturbation theory cannot be trusted. So the energy flux is positive within the valid range of perturbation theory. We further comment on the positivity of the energy flux in the next section.

\section{Normalised Energy Flux}\label{Energy flux}
\label{sec:energy}

In this section, we compute the normalised energy flux sourced by a perturbation produced by single field insertion with momentum $(q,\vec{0})$ with $q>0$ as measured by a calorimeter at null infinity, up to 3 loops. As explained in section \ref{sec:Gaussian}, we first need to compute the norm of the state, which follows from the Wightman 2-point function. 
After computing the Euclidean 2-point function up to 3-loops with usual diagrammatic methods, we perform the Wick rotation as specified in the section \ref{sec:Wick-rot}, finding
\begin{align}\label{2pnt-L-3loop}
\llangle\phi(p)\,\phi(-p)\rrangle_{\3}=&2\pi\,\frac{\delta(p^{\0}-|\vec{p}\,|)}{p^{\0}+|\vec{p}|}+\frac{c_2(d)\,\lambda^2}{(d-4)}\,
\frac{1}{(p^{\0}+|\vec{p}\,|)^{5-d}}\,\frac{d}{dp^{\0}} \left( \frac{\Theta(p^{\0}-|\vec{p}\,|)}{(p^{\0}-|\vec{p}\,|)^{4-d}}\right)\notag\\[7pt]
&\hspace{24mm}+\frac{c_3(d)\,\lambda^3}{(d-4)^2}
\frac{1}{(p^{\0}+|\vec{p}\,|)^{7-\frac{3d}{2}}}\,\frac{d}{dp^{\0}}\,\left(\frac{\Theta(p^{\0}-|\vec{p}\,|)}{(p^{\0}-|\vec{p}\,|)^{6-\frac{3d}{2}}}\right),
\end{align}
with
\begin{align}
c_2(d)&=\frac{1}{12\,(4\pi)^{d-1}}\frac{\Gamma(d/2-1)^3}{\Gamma(d-2)\,\Gamma(3d/2-3)},\\
c_3(d)&=\frac{1}{3\,(4\pi)^{\frac{3d}{2}-1}}\frac{\Gamma(3-d/2)^2\,\Gamma(d/2-1)^5}{\Gamma(d-2)^2\,\Gamma(2d-5)\Gamma(5-d)}.
\end{align}
As explained below \eqref{notation-deltas}, for the case of the 2-loop sunset-type contributions, the above 2 and 3-loop corrections are given in terms of explicit derivatives because for $d\leq4$ they are only well-defined distributions if understood as acting by integration by parts. 

In $d=4-2\epsilon$ dimensions, this 2-point function acquires $1/\epsilon$ and $1/\epsilon^2$ poles, which can be renormalised as in Euclidean signature by introducing field and coupling counterterms, thus yielding a finite Lorentzian 2-point function.

The norm of a momentum eigenstate \eqref{wave-packet2} with timelike momentum $(q,\vec{0})$, $q>0$ has a vanishing tree-level contribution, and we find 
\begin{align}\label{norm-3loop}
\llangle\phi(q)\,\phi(-q)\rrangle_{\3}=
	c_2(d)\,\lambda^2\,\frac{1}{q^{10-2d}}\left(1+\frac{3 \,c_3(d)}{2\,c_2(d)}\frac{\lambda}{(d-4)}\,\,\frac{1}{q^{4-d}}\right).
\end{align}

The 3-loop normalised energy flux finally follows from normalising the 3-loop energy flux \eqref{ANEC-3loop} with the 3-loop norm \eqref{norm-3loop}, and gives 
\begin{equation}
E_{\3}=\frac{\langle \mathcal{E}(q)\rangle_{\3}}{\llangle\phi(q)\phi(-q)\rrangle_{\3}}=\frac{q}{S_{d-2}}.
\end{equation}
An exact and very nontrivial cancellation takes places between the 3-loop corrections of the energy flux and the norm of the state. Therefore, we verify equation \eqref{Energy-measurement-conf}. For scalar states, the positivity of the ANEC trivially follows from the positivity of the norm of the state, given that their proportionality factor is the energy. 

In the case of tensorial states, the positivity of ANEC will give non-trivial bounds. In forthcoming work, we will use the technology developed here to study ANEC on tensorial states in $\lambda \phi^4$ theory.

\section*{Acknowledgements}

We would like to thank Emil Akhmedov, Daniele Dorigoni and Stefan Theisen for discussions. TB is supported by STFC grants ST/P000258/1 and ST/T000759/1. 
LC~is supported by the International Max Planck Research School for Mathematical and Physical Aspects of Gravitation, Cosmology and Quantum Field Theory. HG is supported by the European Research Council (ERC)
under the European Union’s Horizon 2020 research and innovation programme (grant agreement No 740209).

\appendix
\section{Well-definedness of the state norm} \label{appendix}

The expectation value of the energy flux is normalised using the  norm  of the state. In this appendix, we consider the norm of various states and show why further regularisation, such as using a Gaussian factor, may be needed to make the norm well-defined. 

Consider the state \eqref{wave-packet},
 \begin{equation} \label{A1}
|\phi(q)\rangle\equiv\int d^d x\, e^{-i q\, x^{\0}} \,\phi(x)|0\rangle,
 \end{equation}
constructed with a single field insertion on the vacuum. The  norm  of this state has a highly  singular behaviour, \eqref{acc2},
\be
\langle 
\phi(q)| \phi(q) 
\rangle \propto \frac{\delta(q)}{q}.
\ee  
This is peculiar to the state generated by the single insertion of the field: If we consider insertion of higher powers of the field, the situation is qualitatively different. 

For clarity, we consider, in the free theory, the state
\begin{equation}\label{taa}
|{\phi^2( q)}\rangle = \int d^dx \ e^{-iq x^0} \phi^2(x) |0\rangle.
\end{equation}
Using Wick's theorem and discarding tadpole contributions, the norm is given in terms of the Lorentzian \(\llangle \phi^2(q,\vec 0) \phi^2(-q,\vec 0)\rrangle \)  correlator. We  construct it starting from the Euclidean expression 
\begin{equation}\label{tab}
 \llangle      \phi^2(p)\phi^2(-p)  \rrangle_{E,0} = \int \frac{d^d k}{(2\pi)^d} \frac{1}{k^2(k-p)^2}  
 \propto 
 \frac{1}{(p^2)^{2-d/2}}\,,
\end{equation}
which has a non-integer exponent. This technical detail has the profound consequence that in the complex \( p_E = i p^0 \)  plane there is a branchcut rather than a simple pole, resulting in the Wick-rotated expression
\begin{equation}
 \llangle      \phi^2(p)\phi^2(-p)  \rrangle_{0} 
 \propto 
 \frac{1}{|p^2|^{2-d/2}} \Theta[p^0-|\vec p|]\,.
\end{equation}
When evaluated in the momentum configuration \(p = (q,\vec 0)\), it produces \(\Theta[q] \, q^{d-4}\), which is well-defined. It is easy to argue that such behaviour is typical when considering the insertion of \(n>1\) powers of the field. We note that this is similar to the case of a general (interacting) CFT where operators have non-integer dimension, which is why Gaussian factors in the spatial directions can be neglected without problem in \cite{HofMal} and \cite{Hartman:2016lgu}. 

We can therefore trace the origin of the singular behaviour \(\delta(q)/q\)  to the decay of the propagator with an integer power of the momentum, that in our discussion is merely accidental. There are different ways to deal with such an expression: In \eqref{ANEC-tree-Gaussian},  we used a Gaussian factor, but this has the drawback of making the loop calculations unwieldy; or, we could also regulate the behaviour of the free propagator introducing a parameter \(0<\alpha<\frac12\) in the Euclidean expression, \(1/p^2 \to 1/[p^2]^{1+\alpha}  \), but this results in much  more involved branchcut integrals for all momenta in the correlators.

Another option is to start with a different state, such as \eqref{taa}, but in this case many more diagrams have to be evaluated.  

The solution that we come up with, in section \ref{sec:positive}, is to use the state defined in \eqref{A1} but with $q>0$. This means that the norm of the state is only non-zero at higher orders in $\lambda.$

	\bibliographystyle{utphys}
	\bibliography{LambdaPhi4}
	
\end{document}